\newcommand\mass{{\mathrm M}}
\newcommand\charg{{\mathrm Q}}
\newcommand\me{{\mathrm m}_{\mathrm e}}
\newcommand\mplanck{{\mathrm m}_{\mathrm{pl}}}
\newcommand\lpla{{\mathrm l}_{\mathrm{pl}}}
\newcommand\echa{{\mathrm e}}
\newcommand\mua{\mu_{\mathrm a}}
\newcommand\hbac{{\mathrm \hbar}\; {\mathrm c}}

\newcommand\beque{\begin{equation*}}
\newcommand\beq{\begin{equation}}
\newcommand\eeq{\end{equation}}
\newcommand\eeque{\end{equation*}}
\newcommand\beqnl{\begin{eqnarray}}
\newcommand\beqna{\begin{eqnarray*}}
\newcommand\eeqna{\end{eqnarray*}}
\newcommand\eeqnl{\end{eqnarray}}

 \def\NN{\hbox{\sf I\kern-.13em\hbox{N}}}
 \def\HH{\hbox{\sf I\kern-.13em\hbox{H}}}
 \def\DD{\hbox{\sf I\kern-.13em\hbox{D}}}
 \def\RR{\hbox{\sf I\kern-.14em\hbox{R}}}
 \def\CC{\hbox{\sf I\kern-.44em\hbox{C}}}
 \def\ZZ{{\hbox{\sf Z\kern-.43emZ}}}
 \def\QQ{\hbox{\sf C\kern -.48emQ}}
 \def\Cc{\hbox{\sf C\kern -.47em {\raise .48ex \hbox{$\scriptscriptstyle |$}}
   \kern-.5em {\raise .48ex \hbox{$\scriptscriptstyle |$}} }}
 \def\Qq{\hbox{\sf Q\kern -.57em {\raise .48ex \hbox{$\scriptscriptstyle |$}}
   \kern-.55em {\raise .48ex \hbox{$\scriptscriptstyle |$}} }}

\documentstyle[aps,prd,epsfig,preprint]{revtex}
\tightenlines

\begin{document}


\title{Naked Reissner-Nordstr\"{o}m Singularities and 
the Anomalous Magnetic Moment of the Electron Field}

\author{F. Belgiorno\footnote{E-mail address: belgiorno@mi.infn.it}}
\address{Dipartimento di Fisica, Universit\`a di Milano, 20133 
Milano, Italy, and\\
I.N.F.N., sezione di Milano, Italy}

\author{M. Martellini\footnote{E-mail address: martellini@mi.infn.it}}
\address{Dipartimento di Fisica, Universit\`a di Milano, 20133 
Milano, Italy\\ 
I.N.F.N, sezione di Milano, Italy, and\\
Landau Network at ``Centro Volta'', Como, Italy}

\author{M. Baldicchi\footnote{E-mail address: baldicchi@mi.infn.it}}
\address{Dipartimento di Fisica, Universit\`a di Milano, 20133 
Milano, Italy, and\\
I.N.F.N, sezione di Milano, Italy}

\maketitle

\vskip 0.5truecm
\hskip 3truecm Published in Phys. Rev. {\bf D62}, (2000) 084014 - APS

\begin{abstract}
\vskip -0.5truecm
We study the problem of the quantization of the massive 
charged Dirac field on a naked Reissner-Nordstr\"{o}m background. 
We show that the introduction of an anomalous magnetic moment 
for the electron field allows a 
well--defined quantum theory for the one-particle Dirac Hamiltonian, 
because   
no boundary condition on the singularity is required. This means that 
would-be higher order corrections can play an essential role in 
determining physics on the naked Reissner-Nordstr\"{o}m background 
and that a non-perturbative approach is required. 
Moreover, we show that bound states for the Dirac equation are allowed. 
Various aspects of the physical picture emerging from our study
are also discussed, such as 
the possibility to obtain exotic atomic systems, the formation of 
black holes by electronic capture and 
some interesting consistency problems involving quantum gravity.

\vskip -0.5truecm
\end{abstract}




\section{Introduction}
\label{intro}

The behavior of a charged massive Dirac field  
on a naked Reissner-Nordstr\"{o}m background is investigated. It is  
known that the Dirac Hamiltonian in the case of minimal 
coupling 
with the Coulomb classical field of the singularity 
requires the choice of a boundary condition on the 
singularity \cite{cohenp,belgio}, 
and so it is affected by the same problem as the 
free Dirac equation. This problem, which amounts, on a mathematical 
footing, to the fact that the Hamiltonian is not 
essentially self--adjoint,    
makes quantum physics not well-defined on the given background. 
Some qualitative similarities occurring 
with the case of the Dirac equation in 
flat space-time in the presence of a strongly charged point--like 
nucleus are also underlined in \cite{belgio}.\\
Here we  study the problem further on. 
In particular, in order to 
estimate the relevance of would-be higher order quantum 
electrodynamics corrections, in sect. \ref{essauto} we introduce an 
anomalous magnetic moment in the Dirac equation. 
Surprisingly, in the case of the electron field, the 
presence of an anomalous magnetic moment, e.g. of the order of the 
usual flat space-time one is shown to be sufficient 
for ensuring the essential 
self--adjointness of the one-particle Hamiltonian, because no boundary 
condition on the singularity is required. A further 
analogy with the flat space-time case appears, because it is known that 
the introduction of an anomalous magnetic moment 
ensures the essential self--adjointness of the Dirac Hamiltonian in the 
external field of an highly charged point--like nucleus, for any 
value of the atomic number Z. 
There is still a difference, 
associated with the existence, in the naked Reissner-Nordstr\"{o}m case, of 
a lower bound on the absolute value of the anomalous magnetic moment 
which is necessary for the essential self--adjointness of the Hamiltonian. In 
any case, this lower bound is fully satisfied by an anomalous 
magnetic moment order of the flat space-time one.\\
Some qualitative spectral properties are studied in sect. \ref{spectrum}. 
We show that essential spectrum contributions\footnote{The definition of 
essential spectrum and the physical interest of this result 
are discussed in sect. \ref{spectrum}.} from near the 
singularity are excluded and that eigenvalues exist and have to belong 
to the mass-gap. Moreover, the presence of an 
infinite number of eigenvalues is verified.\\
In sect. \ref{atom}  
the possibility to construct a ``naked--Reissner-Nordstr\"{o}m atom" 
is sketched and related consistency problems discussed. In particular, 
the possibility to get 
Reissner-Nordstr\"{o}m black holes by electronic capture is 
qualitatively analyzed. Some puzzling consistency problems 
are enhanced, particularly, situations are sketched in which one 
is forced to  introduce a full quantum gravity formalism.\\ 
The final discussion, sect. \ref{conclusions}, takes into account also the 
Cosmic Censorship Conjecture (CCC).\\ 
In appendixes A and B physical dimensions 
involved in the problem and a further comparison with the case of 
flat space--time are found. In appendix C, some proofs of results presented 
in the main text are given; moreover, for the sake of completeness, 
the enunciates of some of the theorems used are found.

\section{Dirac Hamiltonian with anomalous magnetic moment}
\label{essauto}

In this section we check if the one-particle Hamiltonian 
is well-defined in the sense that no boundary conditions 
are required in order to obtain a self-adjoint operator. 
In other terms, we check if the 
Hamiltonian is essentially self-adjoint, that is,  
if a unique self-adjoint extension and a uniquely determined 
physics occur.  
[In order to give a qualitative idea about the problem of 
defining a self-adjoint extension of an operator, 
we note that the one-particle Hamiltonian 
operator we are going to obtain by variable separation 
is a differential operator which represents a formal differential 
expression in a suitable Hilbert space; 
with this formal expression, according to a 
general theory (see e.g. \cite{weidmann,glazman}), 
are associated the minimal operator and the maximal operator
\footnote{The maximal operator is defined on the 
largest possible domain in the Hilbert space which is mapped into 
the Hilbert space itself. 
The  minimal operator is defined as the restriction of 
the maximal one, such that the adjoint of the minimal operator 
is equal to the maximal operator \cite{weidmann}. See also \cite{glazman}.}. 
The minimal operator is to be suitably extended in order to get a 
self-adjoint operator; it is a basic tool for defining the self-adjoint 
operators which can be associated with the original formal differential 
expression (they 
are the so-called self-adjoint realizations of the formal expression 
and correspond to self-adjoint extensions of the minimal operator)
\footnote{They are formally obtained as self-adjoint 
restrictions of the maximal operator.}. 
Often, the Hamiltonian one writes is meant to be identified with 
the corresponding minimal operator. The essential self-adjointness of the 
minimal operator means that a unique self-adjoint operator can be 
associated with the original formal expression. 
From a physical point of view, a unique self-adjoint extension of 
the (minimal operator associated with the) Hamiltonian means 
that the physics is uniquely defined. In the following, for our aim, 
we can limit ourselves to consider the minimal operator associated 
with our reduced Hamiltonian and to study its self-adjointness 
properties].   
\\ 
We first define the one-particle Hamiltonian for 
Dirac massive particles on the naked Reissner-Nordstr\"{o}m geometry. 
We use natural units ${\mathrm \hbar}={\mathrm c}={\mathrm G}=1$ 
and unrationalized electric units. 
The metric of the naked Reissner-Nordstr\"{o}m manifold 
$(t\in R;\ r\in (0,+\infty);\ \Omega\in S^2)$ is 
\beqnl
ds^2&=& -f(r) dt^2+\frac{1}{f(r)} dr^2 +r^2 d\Omega^2\cr
f(r)&=& 1-\frac{2 \mass}{r}+\frac{\charg^2}{r^2};
\label{regeo}
\eeqnl
$\mass$ is the mass and $\charg$ is the charge, and 
$\charg^2>\mass^2$. The vector potential associated with the 
Reissner-Nordstr\"{o}m solution is $A_{\mu}=(-Q/r,0,0,0)$. 
We choose $\charg>0$. 
The anomalous magnetic moment contribution 
in the Dirac equation is 
proportional to $\sigma^{\mu \nu} F_{\mu \nu}$, and is 
the covariant generalization of the usual flat space-time term \cite{thaller}. 
We will consider explicitly the case of the electron field 
(charge $-\echa$). One gets
\beq
(\gamma^{\mu} D_{\mu}+\me+\frac{1}{2}\; \mua\; 
\sigma^{\mu \nu} F_{\mu \nu})\; \psi=0,
\label{diran}
\eeq
where $\mua$ is the anomalous magnetic moment of the Dirac 
field (see appendix A). 
The spherical symmetry of the problem allows to 
separate the variables 
and to study a reduced problem on a fixed eigenvalue sector of the 
angular momentum operator. For a complete deduction of the 
variable separation see e.g. 
\cite{cohenp,soffel}. We get the following reduced Hamiltonian 
\[ H_{red}=\left[
\begin{array}{cc}
\sqrt{f}\; \me- \frac{\echa\; \charg}{r} &  
-f\; \partial_r +k\; \frac{\sqrt{f}}{r}
+ \mua\; \sqrt{f}\; \frac{\charg}{r^2} \cr
f\; \partial_r +k\; \frac{\sqrt{f}}{r} 
+ \mua\; \sqrt{f}\; \frac{\charg}{r^2} 
& -\sqrt{f}\; \me-\frac{\echa\; \charg}{r}
\end{array} 
\right] 
\]
where  $f(r)$ is the same as in (\ref{regeo}), 
$k=\pm (j+1/2) \in \ZZ-\{0\}$ is the 
angular momentum eigenvalue.  The Hilbert space in which 
$H_{red}$ is formally defined 
is the Hilbert space $L^2 [(0,+\infty), 1/f(r)\; dr]^{2}$ 
of the two-dimensional 
vector functions $\vec{g}\equiv \left(\begin{array}{c}
g_1 \cr
g_2 \end{array}\right)$ such that 
$$
\int_0^{+\infty}\; \frac{dr}{f(r)}\; (|g_1 (r)|^2+|g_2 (r)|^2)<\infty.
$$
As a domain for the minimal operator associated with  
$H_{red}$ we can choose the following subset 
of $L^2 [(0,+\infty), 1/f(r)\; dr]^{2}$: the set    
$C_{0}^{\infty}(0,+\infty)^2$ 
of the two-dimensional vector functions 
$\vec{g}$ whose components are smooth and of compact support 
\cite{weidmann}. 
It is useful to define a new variable $x$ 
as in \cite{belgio} 
\beqna
\frac{dx}{dr}&=&\frac{1}{f(r)}\cr
x&=&r+\mass\ \log(\frac{r^2-2 \mass r+\charg^2}{\charg^2})+
(2 \mass^2-\charg^2) 
\frac{1}{\sqrt{\charg^2-\mass^2}} \arctan(\frac{r-\mass}{\sqrt{\charg^2-
\mass^2}})+C 
\eeqna
and to choose the arbitrary integration constant $C$ in such a way 
that $x\in (0,+\infty)$. The reduced Hamiltonian becomes 
\beq
H_{red}=D_{0}+V(x)
\label{red}
\eeq
where
\[ D_{0}=\left[
\begin{array}{cc}
0 &  - \partial_x \cr 
\partial_x &\ 0\end{array} \right] 
\]
and 
\[ V(r(x))=\left[
\begin{array}{cc}
\sqrt{f}\; \me- \frac{\echa\; \charg}{r} &  +k\; \frac{\sqrt{f}}{r} 
+ \mua\; \sqrt{f}\; \frac{\charg}{r^2} \cr 
+k\; \frac{\sqrt{f}}{r}+ \mua\; \sqrt{f} 
\frac{\charg}{r^2}  
& -\sqrt{f}\; \me-\frac{\echa\; \charg}{r}\end{array} \right]. 
\]
The Hilbert space of interest for the 
Hamiltonian (\ref{red}) is $L^2 [(0,+\infty), dx]^{2}$. 
We have to check if the reduced Hamiltonian is essentially self-adjoint;  
with this aim, we check if the solutions of the equation
\beq
H_{red}\; g=\lambda\; g
\label{eigen}
\eeq
are square integrable in a right neighborhood of $x=0$ and in 
a left neighbourhood of $x=+\infty$. 
The so called Weyl alternative generalized to a system of first order 
ordinary differential equations (\cite{weidmann}, theorem 5.6) 
states that, if the integrability condition in a right neighbourhood of  
$x=0$ is verified for all the solutions corresponding to 
a fixed value of $\lambda \in \CC$, then it is verified 
for every $\lambda \in \CC$ and the so-called 
limit circle case (LCC) is said to occur. 
This occurrence of LCC implies the necessity to 
introduce boundary conditions in order to 
obtain a self-adjoint operator. 
If at least one solution not square 
integrable exists for every $\lambda \in \CC$, then no 
boundary condition is required and 
the limit point case (LPC) is said to be verified. The 
same reasonement is to be applied for $x=+\infty$. 
The Hamiltonian operator is essentially self-adjoint if the LPC 
is verified both at $x=0$ and at infinity 
(cf. \cite{weidmann}, theorem 5.7).\\
It is known \cite{cohenp,belgio} that, if $\mua=0$, 
the reduced Hamiltonian is not essentially 
self--adjoint on the set $C_{0}^{\infty}(0,+\infty)^2$. 
In fact, the 
limit circle case (LCC) at $x=0$ occurs, whereas  
the limit point case (LPC) is verified 
at infinity.\\ 
We show that the introduction of the anomalous 
magnetic moment allows to get the LPC also at $x=0$ for suitable 
values of $\mua$ (the LPC is trivially verified at infinity). 
In our case one gets the 
following system of first order equations in the variable $r$:
\beqna
\partial_r g_{1} +(\frac{k}{\sqrt{f}\; r} 
+ \mua \frac{1}{\sqrt{f}}\; \frac{\charg}{r^2})\; g_{1} + 
\left[- \frac{\me}{\sqrt{f}}+\frac{1}{f} (-\frac{\echa\; \charg}{r}-\lambda) 
\right] g_{2} 
&=&0\cr  
-\partial_r g_{2} +(\frac{k}{\sqrt{f}\; r} 
+ \mua\; \frac{1}{\sqrt{f}}\; \frac{\charg}{r^2}) 
g_{2} + 
\left[ \frac{\me}{\sqrt{f}}+\frac{1}{f}\; (-\frac{\echa\; \charg}{r}-\lambda) 
\right] g_{1} 
&=&0.
\eeqna  
For $r\to 0 \Leftrightarrow x\to 0$ we get the 
following asymptotic expansion for the eigenvalue equation (\ref{eigen}):
\beqna
\partial_r g_{1} +\frac{\mua}{r}\; g_{1}+(\frac{k}{\charg}+
\mua\; \frac{\mass}{\charg^2})\; 
g_{1}&=&O(r)\cr 
\partial_r g_{2} -\frac{\mua}{r}\; g_{2}-(\frac{k}{\charg}+
\mua\; \frac{\mass}{\charg^2})\; 
g_{2}&=&O(r). 
\eeqna
The anomalous magnetic moment contribution is such that 
the coefficients of the asymptotic expansion are no more regular 
near $r=0$. The solutions in a right neighborhood of $r=0$ behave 
as
\beqnl
g_{1}(r)&\sim&a_1\; r^{-\mua}\cr
g_{2}(r)&\sim&a_2\; r^{+\mua}.
\label{nearzero}
\eeqnl
Solutions of (\ref{eigen}) belong 
to $L^2 [(0,R), 1/f(r)\; dr]^{2}$ for $R>0$ if 
\beq
\int_{0}^{R}\; dr\; \frac{1}{f(r)}\; (|g_{1}(r)|^2+|g_{2}(r)|^2)<\infty
\eeq
which means in our case 
\beq
\int_{0}^{R}\; dr\; r^{2\pm 2\; \mua} <\infty.
\label{anom}
\eeq
The above condition implies 
$|\mua|<3/2$. For an explicit evaluation it is necessary to 
resort all the physical dimensions; see appendix A herein. 
If the anomalous magnetic moment value 
is assumed to be the same as in flat space-time, 
for $|\mua|$ in (\ref{anom}) one gets  
$0.00058 \cdot \echa^{\ast}/\me^{\ast} \sim 10^{18}$, where 
$ \echa^{\ast}, \me^{\ast}$ are the lengths associated with the 
electron charge $\echa$ and the electron mass $\me$ respectively;  
then the LPC holds and the reduced Hamiltonian is 
essentially self--adjoint.  It can be noted that the essential 
self--adjointness 
property of the reduced Hamiltonian does not 
depend on the charge $\charg$ and on the mass $\mass$ of the singularity.

\subsection{discussion}

There is a preliminary problem: A perturbative evaluation of 
the anomalous magnetic moment is not available, being the 
free and also the minimally coupled Dirac equation 
on the naked Reissner-Nordstr\"{o}m background 
not well-defined. Nevertheless, 
it is legitimate to consider the anomalous magnetic moment 
as a parameter modeled on the standard QED theory; 
if it is not at least eighteen magnitude 
orders smaller than the standard flat space-time anomalous 
magnetic moment, then the interacting 
theory becomes well-defined. The possibility to have an uniquely 
defined physics {\sl only} for the interacting theory is non-trivial and, 
to some extent, unexpected. We interpret the fact that a 
well--defined physics for the free theory is not available, 
but the interacting theory can avoid this pathologic behavior, 
as the breakdown of the ``perturbative approach'' to the physics.  
In other words, would-be 
higher order dynamical effects in perturbation theory 
actually play a fundamental role in 
determining physics, they determine indeed 
an unique self--adjoint extension of the Hamiltonian.\\ 
According to our proposal, the problem of uniqueness of the 
quantum evolution on a naked 
Reissner-Nordstr\"{o}m manifold becomes a non-trivial dynamical problem 
in a non--perturbative domain.      
A non-perturbative approach is still problematic, 
because one should also verify that no physically 
relevant term has been neglected in the truncation of the effective 
action which gives rise to (\ref{diran}). Nevertheless, 
no matter how limited our exploration of such a domain 
may be, our result opens up a new 
interesting level in the discussion of physics on non-globally 
hyperbolic manifolds.  
A further discussion is found in the conclusions.\\ 

It is remarkable that, to some extent, there can be found an analogy  
with the standard flat space-time Dirac equation in the Coulomb 
external field of an highly charged point--like nucleus. 
In fact, it is known that in flat space-time the anomalous magnetic moment 
solves self--adjointness problems even in the case of an 
heavily charged point--like nucleus \cite{simon}. 
In flat 
space-time the free Dirac Hamiltonian is, of course, essentially 
self--adjoint and the Dirac Hamiltonian 
in an external Coulomb field is essentially self--adjoint too 
as far as ${\mathrm Z}\leq 118$; for bigger Z, 
the essential self-adjointness can be 
restored by introducing the anomalous magnetic moment \cite{simon}
\footnote{For $119\leq {\mathrm Z}<137$ a privileged self--adjoint 
extension can be selected on physical grounds, so that 
the non-trivial  
part of the problem from a physical point of view arises 
for ${\mathrm Z}\geq 137$.}. 
In the naked Reissner-Nordstr\"{o}m case, instead, neither the 
free Dirac Hamiltonian nor the one which is minimally coupled with the 
external Coulomb field of the singularity is essentially 
self--adjoint.\\ 
Moreover, in the case of the Dirac equation in the field of a 
charged point--like nucleus 
it is evident that, as far as the effective coupling of a 
point--like particle 
${\mathrm Z}\; \alpha_{\echa}={\mathrm Z}/137$ 
approaches 1, the perturbative approach looses 
its validity and a non--perturbative approach is necessary. 

In the following,  
some physical properties of our one-particle Hamiltonian 
are discussed.

\section{Spectral properties}
\label{spectrum}

We now study some qualitative spectral properties of the reduced 
Hamiltonian (\ref{red}). It will be found that the essential 
spectrum $\sigma_e (H_{red})$ (defined below) coincides with the complement of 
the interval $(-\me,\me)$, 
and that an infinite number of eigenvalues is confined in the mass-gap.

\subsection{essential spectrum}

The essential spectrum $\sigma_e (B)$ of a self-adjoint operator $B$ consists 
of all points of the spectrum except for isolated eigenvalues of 
finite multiplicity. So it corresponds to the 
union of the continuous spectrum, of the eigenvalues embedded in the 
continuous spectrum or at the edges of the continuous spectrum, 
of the limit points 
for the eigenvalues and of the eigenvalues having infinite 
multiplicity \cite{richtmyer,weidbook} (the latter case cannot 
occur for ordinary differential operators \cite{weidmann}). 
The physical interest is associated 
with the possibility to find, by means of qualitative spectral 
methods, a set which is the complement in the spectrum 
of the set composed by isolated eigenvalues (``bound states''). 
In fact, for any self-adjoint operator $B$ the spectrum 
can be decomposed into the union of two disjoint sets: 
$\sigma (B)=\sigma_e (B)\cup \sigma_d (B)$, where $\sigma_d (B)$ 
is the discrete spectrum, i.e. 
the set containing all the isolated eigenvalues of finite 
multiplicity.\\ 
Let us consider the operators $H_{0}$ and $H_{\infty}$ which are 
defined as the restrictions of $H_{red}$ to the intervals $(0,c]$ and 
$[c,\infty)$, where $c>0$ is arbitrary. 
By using the so called 
decomposition method (\cite{weidmann}, p. 165), 
the essential spectrum of our Hamiltonian 
operator can be decomposed into the union of the essential 
spectra of the operators $H_{0}$ and $H_{\infty}$, in the sense that 
$\sigma_e (H_{red})=\sigma_e (H_{0})\cup \sigma_e (H_{\infty})$   
(see also \cite{belgio}). 
The restriction $H_{\infty}$ of $H_{red}$ 
gives the same essential spectrum 
contribution as the one calculated in \cite{belgio}: 
\beq
\sigma_{e} (H_{\infty})=(-\infty,-\me]\cup [\me,+\infty),
\eeq
as it can be easily verified by using theorems 16.5 and 16.6 of 
\cite{weidmann} ( see appendix C both for the enunciates and 
for their application to our case and  
cf. \cite{bekn} for an application to Kerr-Newman 
black holes). 
The case of $H_{0}$, which is the restriction to 
the right neighborhood of 
the singularity, is a little more involved than in \cite{belgio}, 
because the LPC at the singularity is verified. 
Nevertheless, a 
careful application of theorem 2 appearing in ref. \cite{hintshaw}
allows to obtain the following result: The essential spectrum of the 
Dirac Hamiltonian restricted to a right neighborhood of the 
singularity $r=0$ is empty. We first discuss the physical meaning 
of this result; then we give some more detail. 
The absence of an essential spectrum contribution coming from 
near $r=0$ can be interpreted by means of an analogy with 
standard scattering centers. In fact, avoiding 
essential spectrum contribution from near the center 
amounts to verifying that the one--particle 
scattering problem is well-defined, in the sense that particles 
are not ``captured" for long periods of times near the centers and the 
scattering matrix is unitary. In our case we can analogously say 
that Dirac 
particles don't spend an infinite amount of time near the singularity 
when scattering takes place. See also \cite{horowitz} for the 
case of other time-like singularities.\\
Giving all the details about the cited theorem would require a 
long digression. We limit ourselves to underline that,  
according to the aforementioned theorem, 
if the LPC is verified at $r=0$, in order that 
in $(0,R]$ there can be only a discrete 
spectrum contribution it is sufficient to verify that for 
an arbitrary $R>0$ it holds 
\beq
\int_{0}^{R}\; dr\; \frac{1}{f}\; |k\; \frac{\sqrt{f}}{r} 
+ \mua\; \sqrt{f}\; \frac{\charg}{r^2}|=\infty.
\eeq
In our case the above integral diverges 
because the anomalous magnetic moment gives rise to a term  
which is not integrable in a right neighborhood of $r=0$. 
We refer the interested reader to \cite{hintshaw} for more details. 
Actually, a more naive argument can also be used. 
In the case of a Schr\"{o}dinger--like second order operator 
$\tau$ in $(0,R]$, if the LPC is verified in $r=0$, 
the absence of continuous spectrum for real 
$\lambda\in (\lambda_1,\lambda_2)$ is obtained if the asymptotic 
behavior of the solutions of the differential equation 
$(\tau -\lambda)f=0$ near the origin is such that one is always 
square--integrable. 
In the case of the separated Dirac operator $H_{red}$ the analogous argument 
is found in \cite{weidmann} (theorem 11.7), and  
in our case there is always a square-integrable solution of 
$(H_{red}-\lambda)g=0$ for each $\lambda \in \RR$, as (\ref{nearzero}) 
shows.\\

Finally, note that, as for the analogous equation in flat 
space-time, the interval $(-\me,\me)$ represents a gap in the 
Hamiltonian spectrum between the continuum positive energy 
states and the negative energy ones, and the discrete spectrum 
(isolated eigenvalues) can be located only in $(-\me,\me)$.

\subsection{discrete spectrum}

Here we are interested in the discrete spectrum of the 
one-particle Hamiltonian.    
In the gap $(-\me,\me)$ there is an infinite number 
of discrete eigenvalues. The interested reader 
is referred to appendix C for a proof, which is based 
on theorems given in \cite{hintming}. 
The presence of an infinite number of eigenvalues can be 
considered as a non trivial result (note that 
the proof contained in appendix C 
for the existence of an infinite number of eigenvalues 
holds also when there is no anomalous magnetic moment 
for the electron field). 
In fact, in the case of Reissner-Nordstr\"{o}m black holes 
no isolated eigenvalue is allowed, as it 
is shown in \cite{cohenp,belgio} and in 
\cite{bekn} (in \cite{cohenp} a stronger result is given: no 
eigenvalue exists, no matter if isolated or embedded in the 
continuous spectrum). 
In fact, the presence of the 
black hole horizon does not allow a gap in the essential spectrum of 
the one--particle Dirac Hamiltonian operator \cite{bekn}. 
Then, also from this point of view, 
naked singularities  differ with respect to black holes.

\subsection{purely absolutely continuous spectrum}

We are interested in determining 
if there are eigenvalues embedded into the continuous spectrum. 
Naively, it could be expected that eigenvalues are 
allowed to dive into the continuum as Z increases. A 
careful application of theorem 16.7 of \cite{weidmann} 
[theorem 16.7 of \cite{weidmann} and its application to our 
case are found herein in appendix C] 
shows that the complement of the closed interval $[-\me,\me]$ belongs 
to the purely absolutely continuous spectrum: This means that 
the states with energy in $(-\infty,-\me)\cup (\me,+\infty)$ are 
scattering states with no eigenvalue embedded.  
The physical consequences of this result 
are very interesting:  The eigenvalues have to be confined in the 
mass gap. 
So, contrary 
to the naive expectation, by increasing ${\mathrm Z}$ (Z finite), 
the bound--state energy cannot increase arbitrarily.  
The repulsive nature of the anomalous magnetic 
moment term should be the reason for such a behavior.

\section{A naked--Reissner-Nordstr\"{o}m atom?}
\label{atom}

The existence of stationary states\footnote{For the case of absence 
of anomalous magnetic moment, see \cite{cohenp}.} we 
have shown in the previous section allows us to speculate naively 
about the possibility to dress a 
naked Reissner-Nordstr\"{o}m singularity by means of a 
cloud of electrons, and to obtain, as a consequence, a 
quantum-mechanical object (atomic system).  
In fact, one a priori can fill the bound state energy levels by 
means of electrons and, by pursuing this dressing process,  
the charged singularity can also be neutralized. 
Moreover, one can introduce a sort of ``quantum radius" 
of the singularity, a length scale which appears only at the 
quantum level and corresponds to the Bohr radius for standard atoms
\footnote{The authors are indebted to A.Treves for this 
suggestion.}.\\ 
In the following, we limit ourselves to a qualitative analysis of  
the ``dressing'' of a naked singularity. 
Quantitative evaluations imply very subtle numerical computations, 
because of the non-trivial form of Dirac equation (\ref{diran}) 
in our case.
 
A qualitative picture involves substantially 
two cases.  
In the case of a complete dressing of the singularity  
the space-time metric for a distant observer outside the 
outermost electronic shell (characterized by a radius we will 
call o-radius) is the Schwarzschild one, at least as far as 
multi-pole electromagnetic field contributions associated 
with the electronic shells can be neglected. 
The ``dressed singularity" is 
characterized by a mass order of the original naked singularity 
one (if the total mass of the surrounding electrons is negligible; 
see the discussion below). 
Naively, to this neutral system an effective Schwarzschild 
radius (s-radius in the following) can also be assigned. 
If the dressing is only partial, then the external metric 
becomes a Reissner-Nordstr\"{o}m one but with a reduced 
charge-to-mass ratio with respect to the original naked solution. 
For an exotic atomic system whose ``nucleus" is represented by a naked 
Reissner-Nordstr\"{o}m singularity and whose orbitals are 
filled with standard electrons, an  
electromagnetic spectrum associated with allowed transitions 
between atomic levels is also expected.\\ 
We will also verify that a too naive marriage between general relativity 
(Reissner-Nordstr\"{o}m singularity playing the role of ``nucleus'') 
and quantum mechanical 
orbits 
(electron states surrounding the singularity) 
is not free from ambiguities and possible inconsistencies.\\

We start by making some estimates; with this aim, we restore 
the physical dimensions 
and write the charge-to-mass ratio as
\beq
\frac{\charg^{\ast}}{\mass^{\ast}}=\frac{\sqrt{\alpha_{\mathrm e}}\; 
\frac{\charg}{\echa}}{\frac{\mass}{\mplanck}}=\sqrt{\alpha_{\mathrm e}}\; 
{\mathrm Z}\; 
\frac{\mplanck}{\mass}\simeq 9.35 \cdot 10^{-40}\; {\mathrm Z}\; 
\frac{\mass_s}{\mass},
\eeq
where $\charg^{\ast},\mass^{\ast}$ are 
the lengths associated with Q and M respectively (see also appendix A);  
$\mass_s$ is the Sun mass. 
A naked Reissner-Nordstr\"{o}m singularity is characterized by  
$\charg^{\ast}/\mass^{\ast}\equiv 1+d^2>1$, that is 
\beq
{\mathrm Z}=1.07 \cdot 10^{39}\; \frac{\mass}{\mass_s}\; (1+d^2).
\label{zetel}
\eeq
The parameter $d>0$ points out ``how naked'' the singularity is, 
i.e. how much bigger than one the charge-to-mass ratio is. 
The mass of the singularity being equal, the amount of electrons 
neutralizing the naked singularity is lowest when  
$d^2 \ll 1$. Below we make some estimates for 
${\mathrm Z}$ in the case of small $d$: 
\[
\begin{array}{lcl}
\mass=\mass_s & \Rightarrow & {\mathrm Z}\sim 10^{39}\cr 
\mass=10^{-16} \mass_s & \Rightarrow & {\mathrm Z}\sim 10^{23}\cr 
\mass=\mplanck & \Rightarrow & {\mathrm Z}\sim 12. 
\end{array}
\]
Then, in order to neutralize a naked Reissner-Nordstr\"{o}m singularity 
with a mass order of the Sun mass and with a charge-to-mass ratio only 
slightly bigger that one, at least order of $10^{39}$ electrons 
would be required. We see also that a value of ${\mathrm Z}$ 
order of the standard atomic values is possible only if the mass of the 
singularity is order of the Planck mass. 
For small $d$, it is 
consistent to neglect the electron 
contribution to the total mass of the exotic atomic system: 
In fact, from (\ref{zetel}) one deduces that 
there are about 21 orders of magnitude between the mass $\mass$ and 
the total electron mass contribution ($\mass_s\sim 10^{60} \me$), 
and this means that electron contribution to the mass starts being non 
negligible only if $d^2\sim 10^{20}$. It is then straightforward
to estimate the s-radius of the neutralized system by means of 
the mass $\mass$:
\[
\begin{array}{lcl}
\mass=\mass_s & \Rightarrow & r_s \sim 3\; \hbox{km}\cr 
\mass=10^{-16} \mass_s & \Rightarrow & r_s\sim 300\; \hbox{fm}\cr 
\mass=\mplanck & \Rightarrow & r_s= 2\; \lpla. 
\end{array}
\]

\subsection{dressing and black hole formation problem}

Herein we check if a Reissner-Nordstr\"{o}m naked singularity  
could become a Reissner-Nordstr\"{o}m black hole  
by means of the capture of  N$<$Z electrons; particularly, 
the radius of the electronic shells is compared with the  
Reissner-Nordstr\"{o}m black hole radius $r_+$ associated with 
the dressed solution. (See Fig. 1). A discussion of  
related consistency problems follows. 

%
\begin{figure}[h!]
\setlength{\unitlength}{1.0mm}
\begin{picture}(40,40)
\centerline{\epsfig{figure=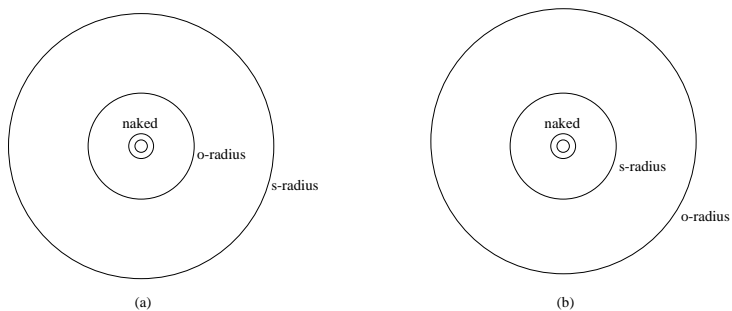,height=4cm,angle=0}}
\end{picture}
\vspace{0.2cm}
\caption{Possible dressings of a naked singularity neutralized 
by electron capture (naive classical picture). 
(a): an effective black hole solution is generated, because the 
electronic shells are within the s-radius; (b): an electronic 
cloud available for external observers is displayed.}
\end{figure}
%
\noindent In general, we write $\mass \equiv 
{\mathrm y}\; \mplanck$, where y$\in (0,+\infty)$ is a real positive number. 
Then 
\beq
\frac{\charg^{\ast}}{\mass^{\ast}}=  
\frac{\mathrm Z\;\echa^{\ast}}{{\mathrm y}\; \lpla}=1+d^2>1;
\eeq
the second equality above fixes the value of y as follows
\beq
{\mathrm y}=\frac{\mathrm Z\;\echa^{\ast}}{(1+d^2)\; \lpla}.
\eeq
When N electrons are captured, from the point of view of an observer 
which is far from the outermost electronic shell, the effective charge 
is $\charg_{\mathrm{eff}}=(\mathrm{ Z-N})\; \echa$, and the effective mass 
is $\mass_{\mathrm{eff}}={\mathrm y}\; \mplanck + {\mathrm N}\; \me$, so that 
\beq
R \equiv \frac{\charg_{\mathrm{eff}}^{\ast}}{\mass_{\mathrm{eff}}^{\ast}}
=\frac{(\mathrm{ Z-N})\; (1+d^2)}{\mathrm{ Z+N}\; (1+d^2)\; 
\frac{\me^{\ast}}{\echa^{\ast}}}.
\eeq
A necessary condition in order to get an horizon is 
$R\leq 1$, which can be obtained for 
\beq 
{\mathrm N}\geq {\mathrm Z}\; 
\frac{d^2}{(1+d^2)\; (1+\frac{\me^{\ast}}{\echa^{\ast}})}
\eeq
(see Fig. 2 for a plot of $N/Z$). Correspondingly, 
the black hole radius would be 
\beq
r_{+}=\mass_{\mathrm{eff}}^{\ast}\; (1+\sqrt{1-R^2}).
\eeq
We choose again to work in the limit of $d\ll 1$, and, in particular, 
as a sample estimate, we impose the condition ${\mathrm Z}\; d^2=1$. 
%
\begin{figure}[h]
\vspace{0.5cm}
  \begin{center}
    \leavevmode
    \setlength{\unitlength}{1.0mm}
    \begin{picture}(62,62)
      \put(10,0){\mbox{\epsfig{file=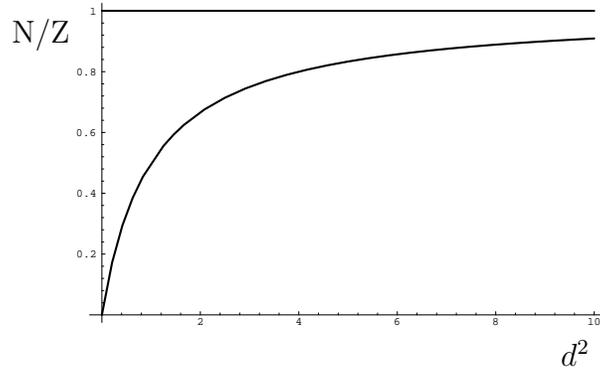,height=23cm}}}
      \put(-10,55){ N/Z }
      \put(63,12){ $ d^2 $ }
    \end{picture}
  \end{center}
\vspace{-1.6cm}
\caption{A plot of the ratio N/Z as a function of $d^2$ is shown. 
For $d^2=9$ the ratio is already order of 0.9.}
\end{figure}
%
\noindent
Then one finds that N$=1$ is enough to obtain $R<1$; 
moreover, one finds ${\mathrm y}\sim 8.54\cdot 10^{-2}\; {\mathrm Z}$, 
$r_{+}\sim {\mathrm y}\; \lpla$.\\ 
Then we consider two cases a), b)  
which appear meaningful.\\ 
Case a): for ${\mathrm Z}=100$ 
the radius is $r_{+}\sim 8.5\; \lpla$ (the mass is 
$\mass = 8.5\; \mplanck$) 
and it is plausible that $r_{+}$ is smaller than the Bohr radius.\\ 
Case b): for ${\mathrm Z}=10^{24}$, 
one gets  
$\mass \sim 10^{15}$\ Kg and 
$r_{+}\sim 10^3$\ fm, and an almost ``atomic'' scale 
appears to be available, to be compared with a huge value of the 
atomic number which would make plausible that the Bohr radius 
is smaller than the estimated $r_{+}$\footnote{It is also plausible that 
it is not necessary to 
approach ${\mathrm Z}=10^{24}$ in order to get $r_{+}>r_{\mathrm{Bohr}}$.}.\\ 
At first sight, the 
second example can allow a picture of transformation of the 
naked singularity into a black hole by means of the capture 
of a single electron, but this conclusion is puzzling:  
A single electron in case b) could be enough to 
induce the appearance of a black hole horizon, in spite of the 
fact that its backreaction is negligible (one has $\me\ll \mass,\;  
\echa \ll \charg$, which should allow for a safe external field 
approximation in the Dirac equation). Moreover,   
in case a), where the backreaction effect of one electron 
is more significant, the electronic capture is not able to transform 
the naked singularity into a black hole. 
A qualitative reason for this paradoxical 
behavior could be that 
in case b) the naked solution 
is much closer to the extremal limit 
$\charg^{\ast}/\mass^{\ast}= 1$ than in case a) ($d^2=10^{-24}$ 
against $d^2=0.01$), so that it should be affected by a much bigger 
instability with respect to electronic capture. 
Even assuming the plausibility of such a picture, 
the mechanism of the generation of the black hole remains unclear.  
However, note that, at the classical level, the transformation of 
naked Reissner-Nordstr\"{o}m singularities into 
Reissner-Nordstr\"{o}m black holes by means of bombardment 
with charged test particles is allowed in \cite{coheng}. 
A further remark is that, after the generation of a black hole, the  
dressing mechanism by means of electronic orbits would stop, 
because no discrete eigenvalue is allowed for a 
Reissner-Nordstr\"{o}m black hole (cf. \cite{bekn};  
an anomalous magnetic moment contribution does not affect the absence of 
discrete spectrum for the electron field on a Reissner-Nordstr\"{o}m 
black hole manifold).

\subsubsection{problems with quantum mechanics} 
 
For a quantum object like 
ours the notion of ``orbit'' is probabilistic and a comparison 
of the expectation value of $r$ (Bohr radius) with the classical 
black hole radius runs the risk of being too naive. In fact,  
qualitatively, the electron field is distributed 
with radial probability density $P_{\echa}(r)$ around the ``naked 
nucleus''. As a consequence, even in the case that 
$r_{+}<r_{\mathrm{Bohr}}$ there can be a significant 
non-zero probability that the electron is within 
the black hole radius $r_{+}$. This implies that 
there can be a significant non-zero probability ${\cal P}$
that the solution is a black hole:
\beq
{\cal P}( \hbox{black\ hole})={\cal P}( \hbox{electron\ between\ 0\ and\ 
}r_{+})\in (0,1).
\eeq 
In other words, the metric seems to be necessarily associated 
with a probability ${\cal P}$ to be a black hole and 
$1-{\cal P}$ to be a naked singularity surrounded by an electron. 
Then, serious self-consistency problems can arise 
if the parameters of the effective dressed solution correspond to  
a black hole solution:  
When ${\cal P}(\hbox{black\ hole})$ 
is significantly different from 0 (or 1), 
the above picture turns out to associate with the metric a 
probabilistic interpretation, and a consistent treatment of the problem
requires a quantum gravity approach. 

\subsection{further consistency considerations}
 
Concerning the radius of the innermost electronic orbits, 
we make some qualitative considerations 
which involve the actual availability 
of the external field approximation for the gravitational 
background. For high ${\mathrm Z}$ 
the Coulomb field interaction could give rise to  
extremely small innermost orbits, and for, say, 
${\mathrm Z}\geq {\mathrm Z}_{0}$ one could find 
an orbit radius smaller than the Planck length, 
in evident conflict with the bound 
on the  minimal length $\lpla$ imposed by quantum gravity. 
In order to be more explicit, let us assume, 
on a purely heuristic footing, that the innermost 
electronic radius scales as $1/{\mathrm Z}$  and satisfies the same 
law as the Bohr radius of an hydrogen-like atom\footnote{For high Z, 
of course, a relativistic approach is necessary for a 
hydrogen-like atom and the non-relativistic formula looses its 
meaning. Herein, the formula is used well beyond its validity 
range, but in the frame of a purely heuristic reasoning.}: 
$r_{\mathrm{Bohr}}= (0.529/{\mathrm Z})\cdot 10^{-10}\; \hbox{m}$. Then, 
for ${\mathrm Z}>10^{25}$ the problem we are discussing surely takes 
place, 
because $r_{\mathrm{Bohr}}<\lpla$ (for ${\mathrm Z}\sim 10^{39}$ one finds 
e.g. $r_{\mathrm{Bohr}} \sim 10^{-50}\; 
\hbox{m}$). If the full problem (i.e. naked singularity 
geometry and anomalous magnetic moment contribution) displays 
an analogous behavior at least for ${\mathrm Z}\geq {\mathrm Z}_{0}$, then 
consistency problems of the semi-classical 
approach for ${\mathrm Z}\geq {\mathrm Z}_{0}$
arise. As a consequence, overcoming the 
problem for boundary conditions on the singularity could be 
insufficient to ensure a full self-consistency of physics at least 
under suitable conditions (e.g. for ${\mathrm Z}\geq {\mathrm Z}_{0}$), 
due to a possible breakdown of the external field approximation 
for the gravitational part of the path-integral. See on this 
topic also the discussion in \cite{damour}.\\
On the other hand, if one introduces {\sl ab initio} 
a box with radius $\sim \lpla$ around the singularity \cite{damour},   
the problem of imposing a boundary condition near the origin 
becomes again unavoidable even in presence of an anomalous 
magnetic moment. A more naive approach consists in assuming that 
the problem is well-posed only when the would-be Bohr radius 
starts being bigger than the Planck length, that is  
only for ${\mathrm Z}\leq {\mathrm Z}_{0}$.\\

Solving the problem of constructing explicitly the exotic 
naked--Reissner-Nord\-str\"{o}m atom is beyond the aim of our work. 
We limit ourselves to note that our naive 
picture of ``dressing'' looks like 
the one in \cite{damour} but there are fundamental differences: 
The charged particles which dress the singularity are not related to the 
Klein paradox and are not in principle due to vacuum instability, 
whose presence on the given background cannot be revealed 
by means of a static approach (see also \cite{belgio}). In our 
picture the electrons are captured from the space region around 
the singularity. Moreover, in our work no boundary condition on the 
singularity is required for the quantized field and a  
discussion about a possible black hole formation appears.

\section{Conclusions}
\label{conclusions}

We have shown that at least in the case of the Dirac field, 
a uniquely defined physics can be retrieved on a naked 
Reissner-Nordstr\"{o}m background in four dimensions, by means 
of the introduction of an anomalous magnetic moment which can also 
be much smaller than in flat space-time. 
A substantial breakdown of the perturbative approach to physics 
is the suggestion we propose for interpreting our result. 
It is remarkable, as a consequence, that the problem of a  
well-posed physics on a naked Reissner-Nordstr\"{o}m background 
can involve non-trivially would-be higher order terms. 
This is verified for the charged massive Dirac field, and 
it would be interesting to 
investigate if any higher order corrections could restore the 
essential self--adjointness also in the case of other fundamental 
fields (e.g. the electromagnetic field or the uncharged 
Dirac particles like the neutrino).\\ 
We can also discuss the relation of our result with the CCC. 
The CCC was formulated with the aim to avoid the indefiniteness 
of physics on non-globally hyperbolic manifolds associated with 
naked curvature singularities. 
Studies involving quantum fields, on the other hand, have shown that 
a well-behaved physics can be recovered for free quantum fields 
on the manifold of a class of naked singularities \cite{horowitz}. 
This allows us to relax the need for the CCC for the aforementioned class. 
We have shown that there is a possibility to relax this need even 
in the case of a Dirac field on a naked Reissner-Nordstr\"{o}m manifold.      
We have yet to underline that our test in a non-perturbative 
domain is interesting but not definitive, 
just because of the substantial lack of a criterion allowing us 
to justify approximations for the effective action calculation 
in a non-perturbative domain, and because of the lack of tools 
allowing to treat a full quantum 
calculation for all the fields (which would avoid  
problems with the external field approximation). Nevertheless, 
our analysis shows that a further level of discussion 
has to be introduced.\\  
In the second part of our work, we have analyzed some aspects 
of the physics associated with our Hamiltonian. 
A spectral analysis of the reduced Hamiltonian has been performed 
and it has been verified that an infinite 
set of eigenvalues is present, contrary to what happens in  
the case of black hole Reissner-Nordstr\"{o}m solutions. Then 
the dressing of the singularity by means of the 
formation of an ``exotic'' atomic system  
and related problems have been discussed.\\

\acknowledgments

The authors wish to thank A.Treves for fruitful discussions and 
suggestions and A.Zaffaroni, R. Ruffini and M.Testa 
for interesting comments. 

\appendix

\section{Dimensions}

We here resort  
all physical dimensions. The function $f(r)=1-2 \mass^{\ast}/r+
(\charg^{\ast})^2/r^2$ is characterized by the lengths 
which are associated with the mass $\mass$ and 
the charge $\charg$ of the Reissner-Nordstr\"{o}m solution respectively:
\beqna
\mass^{\ast}&\equiv& \frac{{\mathrm G}}{{\mathrm c}^2}\; \mass=\lpla\; 
\frac{\mass}{\mplanck}\cr
\charg^{\ast}&\equiv& \sqrt{\frac{{\mathrm G}}{{\mathrm c}^4}}\; \charg=
\sqrt{\frac{\lpla}{\mplanck {\mathrm c}^2}}\; 
\charg=\lpla\; \sqrt{\alpha_{\echa}}\; \frac{\charg}{\echa}.
\eeqna
By posing $\charg={\mathrm Z} \cdot \echa$ one gets 
$\charg^{\ast}=\lpla\; \sqrt{\alpha_{\echa}}\; {\mathrm Z}$ and 
\beq
\frac{\charg^{\ast}}{\mass^{\ast}}=\frac{\sqrt{\alpha_{\echa}}\;
\frac{\charg}{\echa}}{\frac{\mass}{\mplanck}}=\sqrt{\alpha_{\echa}}\; 
{\mathrm Z}\; 
\frac{\mplanck}{\mass}.
\eeq 
It is useful to recall that in the case of the electron one has 
\beqna
\echa^{\ast}&=&8.54\cdot 10^{-2}\; \lpla\\
\me^{\ast}&=&4.18\cdot 10^{-23}\; \lpla\\
\alpha_{\mathrm e}&=&(\frac{\echa^{\ast}}{\lpla})^2.
\eeqna
The anomalous magnetic moment of the electron is given by
\beq
\mua \equiv - a\; \mu_{\mathrm{Bohr}}
\eeq
where $a$ is a dimensionless constant\footnote{The first 
perturbative order in QED in flat space-time 
gives $a=\alpha_{\mathrm e}/(2\pi)$.} and 
$\mu_{\mathrm{Bohr}}$ is the standard Bohr magneton:
\beq
\mu_{\mathrm{Bohr}}=\frac{\echa\; {\mathrm \hbar}}{2\; \me\; {\mathrm c}}.
\eeq
We also write the reduced Hamiltonian as follows 
\[ H_{red}=\left[
\begin{array}{ll}
A&  C_{-}\cr 
C_{+}& B\end{array} 
\right],
\]
where the physical dimensions in each entry are resorted
\beqna
A\;&=&+\sqrt{f}\; (\me\; {\mathrm c}^2)-{\mathrm Z}\; (\alpha_{\echa}\; 
\hbac)\; \frac{1}{r}\cr
B\;&=&-\sqrt{f}\; (\me\; {\mathrm c}^2)-{\mathrm Z}\; (\alpha_{\echa}\; 
\hbac)\; \frac{1}{r}\cr
C_{+}&=&+(\hbac)\; f\; \partial_r+ 
k\; (\hbac)\; \sqrt{f}\; \frac{1}{r}-a\; 
\frac{(\hbac)^2}{2\; \me\; {\mathrm c}^2}\; ({\mathrm Z}\; \alpha_{\echa})\;
\sqrt{f}\; \frac{1}{r^2}\cr  
C_{-}&=&-(\hbac)\; f\; \partial_r + 
k\; (\hbac)\; \sqrt{f}\; \frac{1}{r}-a\; 
\frac{(\hbac)^2}{2\; \me\; {\mathrm c}^2}\; ({\mathrm Z}\; \alpha_{\echa})\;
\sqrt{f}\; \frac{1}{r^2}.
\eeqna
The asymptotic expansion of the eigenvalue equation for $r\to 0$ 
is (each term is divided by $({\mathrm \hbar}\; {\mathrm c})$ so that 
it has dimensions of the inverse of a length)
\beqna
\partial_r g_{1} +\frac{\mua\; \charg}{\charg^{\ast}\; 
{\mathrm \hbar}\; {\mathrm c}} 
\; \frac{1}{r}\; g_{1}&=&O(1)\cr 
\partial_r g_{2} -\frac{\mua\; \charg}{\charg^{\ast}\; 
{\mathrm \hbar}\; {\mathrm c}} 
\; \frac{1}{r}\; g_{2}&=&O(1). 
\eeqna
We are interested in the dimensionless ratio 
$\frac{|\mua|\; \charg}{\charg^{\ast}\; {\mathrm \hbar}\; 
{\mathrm c}}$ which corresponds to the absolute value of the 
$\mua$ appearing in (\ref{anom}) 
\beq
\frac{|\mua|\; \charg}{\charg^{\ast}\; {\mathrm \hbar}\; 
{\mathrm c}}=\frac{a\; \mu_{\mathrm{ Bohr}}\; \echa}
{\echa^{\ast}\; {\mathrm \hbar}\; {\mathrm c}};
\eeq
then
\beqna
\frac{a\; \echa^2}{\echa^{\ast}\; 2\; \me\; {\mathrm c}^2}&=&
\frac{a\; {\mathrm \hbar}\; \alpha_{\mathrm e}}{\echa^{\ast}\; 2\; 
\me\; {\mathrm c}}
=\frac{a\; \lpla\; \mplanck\; \alpha_{\mathrm e}}{\echa^{\ast}\; 2\; \me}=
\frac{a\; \mplanck\; {\echa^{\ast}}^2}{\echa^{\ast}\; 2\; \me\; \lpla}=
\frac{a}{2}\; \frac{\echa^{\ast}} {\me^{\ast}}=1.18\cdot 10^{18}.
\eeqna
So one gets that $a\; \echa^{\ast}/\me^{\ast}\gg 1$ if the 
value of $a$ is not much smaller than the flat space-time one.\\
We list below the values of some factors appearing in our equation 
(for the anomalous magnetic moment the flat space-time value is assumed):
\beqna
\me\; {\mathrm c}^2 &=& 0.510999\; \mathrm{MeV} \cr 
\alpha_{\echa}&=& 1/137.035989 \cr
\hbac &=& 197.327053\; \mathrm{ MeV\; fm} \cr
\frac{(\hbac)^2}{2\; \me\; {\mathrm c}^2}\; \alpha_{\echa}&=& 278.02803\; 
\mathrm{ MeV\; (fm)^2} \cr
\charg^{\ast}&=& 1.38050219 \cdot 10^{-21} \cdot 
{\mathrm Z}\; \mathrm{fm}\cr
a&=&0.001159.
\eeqna

\section{Comparison with flat space-time}

The asymptotic expansions of the potential $V(r(x))$ as $x\to 0$ 
and $x\to +\infty$ is useful for a comparison with the Dirac 
equation in flat space-time $f=1$. We note that 
\[
x = \frac{r^3}{3\; \charg^2}+O(r^4) \quad \hbox{for}\ r\to 0
\]
and 
\[
x = r + 2\; \mass\; \log(r) +O(1) \quad \hbox{for}\ r\to +\infty
\]
in such a way that $r \sim (3\; \charg^2)^{1/3} x^{1/3}$ 
and $r\sim x$ respectively. Near the singularity one gets 
(only the leading order of each entry is displayed)
\[ V(r(x))\sim \left[
\begin{array}{cc}
(\me-\echa) \left( \frac{\charg}{3} \right)^{\frac{1}{3}} x^{-\frac{1}{3}}
&  \frac{\mua}{3}\; x^{-1}\cr 
\frac{\mua}{3}\; x^{-1}
& (-\me-\echa) \left( \frac{\charg}{3} \right)^{\frac{1}{3}} x^{-\frac{1}{3}}
\end{array} \right], 
\]
and near infinity 
\[ V(r(x))\sim \left[
\begin{array}{cc}
\me+(-\me\; \mass-\echa\; \charg)\; x^{-1} 
&  k\; x^{-1}\cr 
k\; x^{-1} 
& -\me +(\me\; \mass -\echa\; \charg)\; x^{-1}
\end{array} \right]. 
\]
The difference with respect to the flat space-time Hamiltonian is 
mostly evident near the origin, but also near infinity there are 
sub-leading corrections to the behavior of the Dirac Hamiltonian 
in flat space-time.

\section{Theorems on Dirac systems and some proofs}

We list below the enunciates of some theorems we refer to in our paper. 
A Dirac operator of the form 
\[ H=\left[
\begin{array}{cc}
0 &  \partial_x\cr 
-\partial_x &\ 0\end{array} \right]+P(x) 
\]
defined on $I=(a,b)$ will be considered; the potential  
\[ P(x)\equiv \left[
\begin{array}{cc}
p_1(x) &  p_{12}(x)\cr 
p_{12}(x) & p_2(x)\end{array} \right] 
\]
is real symmetric, $|P(x)|$ is locally integrable, 
$p_1(x),p_2(x),p_{12}(x)$ are real functions locally 
integrable \cite{weidz}. 
$|\cdot|$ stays for a norm in $\CC^{2 \times 2}$ 
(e.g. the Euclidean norm for matrices; see below).\\
\\
In order to work with an operator having the form required by theorems 
appearing in \cite{weidmann}, we introduce the unitary matrix 
\[ T\equiv \left[
\begin{array}{cc}
0 &  1\cr 
1 &  0\end{array} \right] 
\]
and an operator 
$H_{\ast}\equiv T\; H_{red}\; T^{\dagger}$ which is 
unitarily equivalent to $H_{red}$ (so it has the same spectrum 
and the same spectral properties as $H_{red}$) and 
matches the required form. In particular, 
we have 
\[
H_{\ast}=\left[
\begin{array}{cc}
0 &  \partial_x\cr 
-\partial_x &\ 0\end{array} \right]+P(x) 
\]
where 
\[P(x)\equiv T\; V(r(x))\; T^{\dagger}=\left[
\begin{array}{cc}
-\sqrt{f}\; \me- \frac{\echa\; \charg}{r} &  +k\; \frac{\sqrt{f}}{r} 
+ \mua\; \sqrt{f}\; \frac{\charg}{r^2} \cr 
+k\; \frac{\sqrt{f}}{r}+ \mua\; \sqrt{f} 
\frac{\charg}{r^2}  
& \sqrt{f}\; \me-\frac{\echa\; \charg}{r}\end{array} \right]. 
\]
\\

{\sl Theorem 16.5 of \cite{weidmann}}:\\
Assume that $H$ is regular at $a$\footnote{``Regular at $a$'' (where 
$a$ is finite) means that the assumptions on the coefficients of the 
differential expression $H$ are satisfied in $[a,b)$ instead of 
in $(a,b)$ \cite{weidmann}.} and that $b=+\infty$. If $P(x)\to P_0$ for
$x\to +\infty$ and $\mu_{-}\leq \mu_{+}$ are the eigenvalues of $P_0$, 
then for every self-adjoint extension $H_1$ of $H$ it holds 
$\sigma_e (H_1)\cap (\mu_{-},\mu_{+})=\emptyset$.\\
\\
This theorem is applied to $T\; H_{\infty}\; T^{\dagger}$. 
In our case, $a=c$ and the above operator is regular at $c$; moreover, 
it is easy to show, by taking the limit $\lim_{x\to +\infty}\; P(x)$, 
that $\mu_{-}=-\me$ and $\mu_{+}=\me$. As a consequence of 
the above theorem, $\sigma_e (H_{\infty})\cap (-\me,\me)=\emptyset$.\\ 

{\sl Theorem 16.6 of \cite{weidmann}}:\\ 
Assume that $b=+\infty$ and let 
$\mu_{-}\leq \mu_{+}$ be the eigenvalues of $P_0$ defined as above. 
If for some $d\in(a,+\infty)$ 
\[
\lim_{x\to +\infty} \frac{1}{x}\; \int_d^x\; dt\; |P(t)-P_0|=0
\]
then for every self-adjoint extension $H_1$ of $H$ it holds 
$\sigma_e (H_1)\supset \hbox{complement\ of}\ (\mu_{-},\mu_{+})$.\\ 
\\
This theorem is again applied to $T\; H_{\infty}\; T^{\dagger}$. We can choose 
$d=c$. The Euclidean norm for $P(x)$ is defined as
$$
|P(x)|=\sqrt{|p_1(x)|^2+|p_2(x)|^2+2\; |p_{12}(x)|^2}.
$$
In our case, $|P(x)-P_0|$ is order of $(1/x)$ 
as $x\to +\infty$;   
then $\int_d^x\; dt\; |P(t)-P_0|$ diverges and, by applying  
L'Hospital's rule to $1/x\; \int_d^x\; dt\; |P(t)-P_0|$ one 
finds that the above limit is zero. Theorem 16.6 allows us to 
conclude that $\sigma_e (H_{\infty})\supset (-\infty,-\me]\cup 
[\me,+\infty)$. This result and the above one imply that 
$\sigma_e (H_{\infty})= (-\infty,-\me]\cup [\me,+\infty)$.\\

{\sl Theorem 16.7 of \cite{weidmann}} (see also \cite{weidac}):\\
Consider $H$ satisfying the LPC at $b=+\infty$ 
(LPC or LCC at $a$). Assume that $P(x)$ 
can be decomposed for some $c\in (a,+\infty)$ as follows:
\beqna
&&P(x)=P_{1}(x)+P_{2}(x),\cr
&&|P_{1}(x)|\in L_{1}(c,+\infty),\cr
&&P_{2}(x)\in BV([c,+\infty)),\cr
&&\lim_{x\to +\infty}\; P_{2}(x)= \left[
\begin{array}{cc}
\mu_{1}& 0\cr 
0& \mu_2 \end{array} \right]  
\eeqna 
with $\mu_{1}<\mu_{2}$. 
Then, each self--adjoint extension of $H$ 
has purely absolutely continuous spectrum in the complement 
of $[\mu_{1},\mu_{2}]$.\\
Cf. also \cite{thaller}, theorem 4.18.\\  
In the theorem above, 
$BV([c,+\infty))$ represents the space of 
the functions of bounded variation on the interval $[c,+\infty)$. 
We recall that $f\in BV([c,+\infty))$ means that, for any partition 
$\Pi: c=x_0<x_1<\ldots<x_n=b$ of the interval $[c,b]$, where $c<b<+\infty$, 
the variation   
$$ 
V_c^b (f)\equiv \sup_{\Pi}\; \sum_{k=0}^{n}\; |f(x_k)-f(x_{k-1})| 
$$ 
is finite, and, moreover, 
$$
\lim_{b\to +\infty}\; V_c^b(f)\equiv V_c^{+\infty}(f)
$$ 
exists and is finite.\\
\\
We apply this theorem to $T\; H_{red}\; T^{\dagger}$. 
This Dirac operator satisfies the  
hypotheses of the cited theorem. In fact, in the interval 
$[c,+\infty)$, where $c>0$, 
each component of $P(x)$ is 
smooth and has derivative belonging to $L_1 ([c,+\infty))$, so 
that $P(x)\in BV([c,+\infty))$ 
[note that the anomalous term could as well belong to $P_1(x)$]. 
This follows from the 
fact that, in general, if a function 
$f$ is e.g. continuously differentiable  
and its derivative $f^{\prime}$ belongs to $L_1 ([c,+\infty))$, 
then 
$$
\sum_{k=0}^{n}\; |f(x_k)-f(x_{k-1})|=\sum_{k=0}^{n}\; |\int_{x_{k-1}}^{x_k}\; 
dt\; f^{\prime}(t)|\leq \sum_{k=0}^{n}\; \int_{x_{k-1}}^{x_k}\; 
dt\; |f^{\prime}(t)|=\int_c^b\; dt\; |f^{\prime}(t)|,
$$
and the condition $f^{\prime}\in L_1 ([c,+\infty))$ allows 
to get the desired result. 
Moreover, 
\[\lim_{x\to +\infty}\; P(x)= \left[
\begin{array}{cc}
-\me & 0\cr 
0& \me \end{array} \right].
\]
Then our operator $H_{red}$ 
has a purely absolutely continuous spectrum in the complement of the 
closed interval $[-\me,\me]$.\\   
Note also that this holds also for the flat space-time case [where the 
anomalous contribution is monotone and bounded in $[c,+\infty), c>0$ 
(then it is of bounded variation) and 
is also a term which can belong to $P_1(x)$].

\subsection{discrete spectrum}

In order to verify that an infinite number of eigenvalues is contained in the 
mass-gap of our one-particle Hamiltonian, we use theorem 
2.3 of \cite{hintming}. Some preliminar definitions are given 
below.\\ 
One considers for $x\in (0,+\infty)$ an operator $L$ of the form 
\beq
L\; y\equiv J\; \left(y' - S\; y \right)
\eeq
where
\[ J= \left(
\begin{array}{cc}
0 &  -1 \cr 
1 &\  0\end{array} \right),
\]
and
\[ S= \left(
\begin{array}{cc}
p(x) &  c_1+V_1(x) \cr 
c_2-V_2(x) &\ -p(x)\end{array} \right), 
\]
where $c_1,c_2$ are positive numbers and $p(x),V_1(x),V_2(x)$ are 
real, locally integrable functions \cite{hintming}.\\
We introduce also a non-trivial linear functional $G[.]$ defined on 
real 2$\times$2 matrices $B$ by $G[B]=\langle B\; u, u\rangle$, 
where $u$ is a non 
null 2-vector and $\langle , \rangle$ is the inner product in 
$\RR^2$. $G[B]$ so defined is a positive functional according 
to the definition of \cite{hintming}. Let $I$ be the identity matrix and 
let $P$ be the matrix 
\[ P\equiv J\; S= \left(
\begin{array}{cc}
-c_2+V_2(x) & p(x)  \cr 
p(x) &\ c_1+V_1(x) \end{array} \right). 
\]
Theorem 2.3 of \cite{hintming} is:\\
Let $h>0$, G be a non-trivial positive linear functional and 
assume P locally absolutely continuous. Then, for any 
self--adjoint extension $L_1$ of $L$ the set 
$\sigma (L_1)\cap 
(-h,h)$ is infinite if the scalar differential equation, 
$$
-G[I]\; z''+G[P^2-h^2\; I + (P'\; J-J\; P')/2]\; z=0
$$
is oscillatory\footnote{``Oscillatory at infinity" means that in a 
left neighborhood 
$(b,+\infty)$, with $b>0$, of $+\infty$ all the solutions of the 
above second order scalar differential equation admit 
infinitely many zeroes in $(b,+\infty)$ \cite{hintming}. 
An analogous definition holds for ``oscillatory at $0$".}  
either at 0 or at $+\infty$.\\

We verify 
that our Hamiltonian implements the conditions given in theorem 2.3  
of \cite{hintming}. 
In our case we have 
\beqna
c_1&=&c_2=\me\cr
V_1(x)&=&(\sqrt{f}-1)\; \me + \frac{\echa\; \charg}{r}\cr 
V_2(x)&=&-(\sqrt{f}-1)\; \me + \frac{\echa\; \charg}{r}\cr 
p(x)&=&-(\frac{k\; \sqrt{f}}{r}+\mua\; \sqrt{f}\; 
\frac{\charg}{r^2}).
\eeqna 
As a consequence,  in order to 
verify if the spectrum of the self-adjoint extension $L_1$ of 
$L$ has an infinite number of eigenvalues in $(-\me,\me)$ it is 
sufficient to verify that the following scalar differential 
equation 
\beq
-G[I]\; z''+G[P^2-\me^2\; I + (P'\; J-J\; P')/2]\; z=0
\label{oscill}
\eeq
(where $P\equiv J\; S $)  
has an oscillatory behavior either at $0$ or at 
$+\infty$.
Note that in 
our case the self--adjoint extension of the 
reduced Hamiltonian is unique. 
We choose $u_{+}=(1,0)^T$ and also $u_{-}=(0,1)^T$. 
Then one obtains a scalar equation in the form
\beq
-z''+\Gamma_{\pm}(x)\; z=0
\label{scalar}
\eeq
where $\Gamma_{\pm}$ is relative to the choice of the vector 
$u_{\pm}$. One has 
\beqna
\Gamma_{+}(x)&=& V_2^2(x)-2 \me\; V_2(x)+p^2 (x)+p'(x)\cr
&&\cr
\Gamma_{-}(x)&=& V_1^2(x)+2 \me\; V_1(x)+p^2 (x)-p'(x)
\eeqna
and asymptotically for $x\to +\infty$ it holds 
\beq
\Gamma_{\pm}(x) \sim \frac{- 2\; \me\; (\me\; \mass \pm \echa\; \charg)}{x}.
\eeq
One can use corollary 37, p.1463,
of \cite{ds} for the scalar equation (\ref{scalar}):  
If the limit 
$\lim_{x\to +\infty} x^2 \Gamma_{\pm}(x)=\lim_{r\to +\infty} r^2 
\Gamma_{\pm}(r)<-1/4$, then the equation is 
oscillatory near $+\infty$. In our case $\charg>0$ and $x^2 
\Gamma_{+}(x)\to -\infty$ as $x\to +\infty$ (if $\charg<0$ then $x^2 
\Gamma_{-}(x)\to -\infty$), that is, the behavior is oscillatory. 
Cf. also the examples in \cite{hintming}. Then 
\beq
\sigma (H_{red}) \cap (-\me, \me)=\hbox{infinite set}.
\eeq

\end{document}